\documentstyle[aps,preprint,prb]{revtex}
\begin{document}
\title{\bf PAIRING IN THE BOGOLIUBOV-de GENNES EQUATIONS}
\author{Yong-Jihn Kim$^{\dagger}$}
\address{Department of Physics, Purdue University, West Lafayette, Indiana 47907}
\maketitle
\def\hb{\hfill\break}
\begin{abstract}
 It is shown that the Bogoliubov-de Gennes equations
 pair the electrons in states which are linear combinations of the
 normal states. Accordingly, the BCS-like reduction procedure is required 
 to choose a correct pairing. For a homogeneous system, we point out that 
 the kernel of the self-consistency equation derived from the 
 Bogoliubov-de Gennes equations needs to be constrained by the BCS pairing 
 condition.  In the presence of ordinary impurities, on the other hand, the 
 Bogoliubov-de Gennes equations should be supplemented by Anderson's pairing 
 condition to obtain the correct vacuum state by the corresponding unitary 
 transformation.  This results in localization correction to the 
 phonon-mediated interaction. 
\end{abstract}
\vskip 1pc
\noindent
$^{\dagger}$ Present address: Department of Physics, Korea Advanced 
Institute of Science 

and Technology, Taejon 305-701, Korea
\vfill\eject

\leftline {\bf 1. Introduction} 
\vskip 1pc 

Recently it was shown$^{1,2}$ that the Abrikosov and Gor'kov's$^{3,4}$ 
Green's function theory of impure superconductors is in serious conflict with 
Anderson's theorem.$^{5}$
(Strong coupling theory of impure superconductors was discussed 
elsewhere.$^{6}$) 
For magnetic impurity effects, Kim and Overhauser$^{7}$ (KO) proposed a
BCS type theory with different predictions from the 
pair-breaking theory of Abrikosov and Gor'kov.$^{3}$ In fact,
there have been several experiments$^{8-12}$ which 
agree with KO's predictions.   
It was also pointed out$^{13}$ that the failure of Green's function theory 
comes from the intrinsic pairing problem in Gor'kov's 
formalism.$^{14}$
The self-consistency equation needs to be supplemented by a pairing 
condition derived from
the physical constraint of the Anomalous Green's function.
The resulting equation is nothing but another form of the BCS
gap equation.
Then the discrepancy disappears. 

In this letter we show that the same pairing constraint is required 
in the Bogoliubov-de Gennes (BdG) equations.$^{15}$ 
The equations are obtained by a real space version
of the Bogoliubov-Valatin (BV) transformation,$^{16, 17}$ which is a 
generalization$^{18,19}$ of the BV transformation.
In this case, the pairing constraint is determined to find a 
correct vacuum state by the corresponding unitary transformation. 
The resulting pairing condition is the same as that obtained from the 
physical constraint of the Anomalous Green's function.$^{13}$ 

For a homogeneous system, choosing a constant pair potential gives rise to
the BCS pairing. However, the kernel of the self-consistency equation
has not been fixed by the BCS pairing condition. In the presence of ordinary
impurities, Anderson's pairing between time-reversed states is not obtained 
from the BdG equations  because the pair potential depends
on the position. 
Alternatively, pairing occurs between the states which are the linear
combination of the scattered states. Consequently, the BdG
equations predict incorrectly that $T_{c}$ doesn't
change even if the scattered states are localized. Note that
the linear combination of localized states becomes extended one.
Magnetic impurities will be considered elsewhere.

\vskip 1pc
\leftline{\bf 2. Homogeneous System}
\vskip 1pc
\leftline{\bf 2.1 Bogoliubov-de Gennes Equations}
\vskip 1pc

Let's consider a homogeneous system. We follow first de Gennes' derivation.$^{15}$  
The Hamiltonian is given
\begin{eqnarray}
H =  \int d{\bf r}\sum_{\alpha}\Psi^{\dagger}({\bf r}\alpha)[{\vec {p}^{2}\over 2m}]
\Psi({\bf r}\alpha) - {1\over 2}V\int d{\bf r}\sum_{\alpha\beta}\Psi^{\dagger}({\bf r}\alpha)
\Psi^{\dagger}({\bf r}\beta)\Psi({\bf r}\beta)\Psi({\bf r}\alpha), 
\end{eqnarray}
where the field operator $\Psi({\bf r}\alpha)$ is expanded by 
plane wave basis set $\phi_{\vec k}({\bf r}) = e^{i\vec {k}\cdot \bf r}$, that is,
\begin{eqnarray}
\Psi({\bf r}\alpha) = \sum_{\vec k} \phi_{\vec {k}}({\bf r}) a_{\vec {k}\alpha},
\end{eqnarray}
where $a_{\vec {k}\alpha}$ is a destruction operator for an electron with spin 
$\alpha$.
Using the Gor'kov's factorization$^{14}$, we may get an effective Hamiltonian of the
form
\begin{eqnarray}
H_{eff} =  \int d{\bf r}\{\sum_{\alpha}\Psi^{\dagger}({\bf r}\alpha)[{\vec {p}^{2}\over 2m}]
\Psi({\bf r}\alpha) + \Delta({\bf r}) \Psi^{\dagger}({\bf r}\uparrow) \Psi^{\dagger}({\bf r}\downarrow)
+ \Delta^{*}({\bf r}) \Psi({\bf r}\downarrow) \Psi({\bf r}\uparrow)\},
\end{eqnarray}
where 
\begin{eqnarray}
\Delta({\bf r}) = -V<\Psi({\bf r}\downarrow)\Psi({\bf r}\uparrow)>.
\end{eqnarray}
$\Delta({\bf r})$ is called by the  pair potential at the position $\bf r$.

In order to find the eigenstates and corresponding energies, we perform a 
unitary transformation 
\begin{eqnarray}
\Psi({\bf r}\uparrow) &=& \sum_{n}(\gamma_{n\uparrow}u_{n}({\bf r}) - 
\gamma^{\dagger}_{n\downarrow}v^{*}_{n}({\bf r})), \nonumber\\
\Psi({\bf r}\downarrow) &=& \sum_{n}(\gamma_{n\downarrow}u_{n}({\bf r}) + 
\gamma^{\dagger}_{n\uparrow}v^{*}_{n}({\bf r})), 
\end{eqnarray}
where the $\gamma$ and $\gamma^{\dagger}$ are quasiparticle operators satisfying the fermion
commutation relations
\begin{eqnarray}
\{\gamma_{n\alpha}, \gamma^{\dagger}_{m\beta}\} &=& \delta_{mn}\delta_{\alpha\beta},
\nonumber\\
\{\gamma_{n\alpha}, \gamma_{m\beta}\} &=& 0. 
\end{eqnarray}
By the transformation (5), the effective Hamiltonian may be diagonalized, that is,
\begin{eqnarray}
H_{eff} = E_{g} + \sum_{n, \alpha}\epsilon_{n}\gamma^{\dagger}_{n\alpha}\gamma_{n\alpha}, 
\end{eqnarray}
where $E_{g}$ is the ground state energy of $H_{eff}$ and $\epsilon_{n}$ is the energy of the excitation $n$.
From the condition (7), we obtain the well-known Bogoliubov-de Gennes equations:
\begin{eqnarray}
\epsilon u({\bf r}) &=& H_{e}u({\bf r}) + \Delta({\bf r})v({\bf r}), \nonumber\\ 
\epsilon v({\bf r}) &=& -H_{e}^{*}v({\bf r}) + \Delta^{*}({\bf r})u({\bf r}), 
\end{eqnarray}
where $H_{e} = {\vec {p}^{2}\over 2m}$.
These equations can be written compactly in a matrix form
\begin{eqnarray}
\epsilon\pmatrix{u\cr v\cr} = \left(\matrix{ \matrix H_{e} & \Delta({\bf r}) \cr
                            \Delta^{*}({\bf r}) & -H^{*}_{e} \cr} \right) \pmatrix{u\cr v\cr}.   
\end{eqnarray}
Notice that different eigenfunctions are orthogonal and $\pmatrix{u \cr v\cr}$ is
orthogonal to $\pmatrix{-v^{*}\cr u^{*}\cr}$. 
 
Substituting Eq. (5) into Eq. (4) we find
\begin{eqnarray}
\Delta({\bf r}) = V\sum_{n}v^{*}_{n}({\bf r})u_{n}({\bf r})(1-2f_{n}), 
\end{eqnarray}
where $f_{n} = {1\over {\rm exp}(\beta \epsilon_{n}) +1 }. $
Eq. (10) is the self-consistency equation for the pair potential.

\vskip 1pc
\leftline{\bf 2.2 Vacuum State}
\vskip 1pc

Now we analyze the derivation.
It is important to note that the unitary transformation (5) is in fact 
a generalization$^{18,19}$ of the BV
transformation,
\begin{eqnarray}
a_{\vec {k}\uparrow} &=& u_{k}\gamma_{\vec {k}\uparrow} + v_{k}\gamma^{\dagger}_{-\vec {k}\downarrow}, \nonumber\\
 a_{\vec {k}\downarrow} &=& u_{k}\gamma_{\vec {k}\downarrow} - v_{k}\gamma^{\dagger}_{-\vec {k}\uparrow},
\end{eqnarray}
where 
\begin{eqnarray}
u_{k}^{2} + v_{k}^{2} = 1.
\end{eqnarray}
Inserting Eq. (2) into Eq. (5) we obtain
\begin{eqnarray}
 \sum_{\vec {k}'}\phi_{\vec {k}'}({\bf r})a_{\vec {k}'\uparrow} &=& \sum_{n}(\gamma_{n\uparrow}u_{n}({\bf r}) - 
\gamma^{\dagger}_{n\downarrow}v^{*}_{n}({\bf r})), \nonumber\\
 \sum_{\vec {k}'}\phi_{\vec {k}'}({\bf r})a_{\vec {k}'\downarrow} &=& \sum_{n}(\gamma_{n\downarrow}u_{n}({\bf r}) + 
\gamma^{\dagger}_{n\uparrow}v^{*}_{n}({\bf r})). 
\end{eqnarray}
If we multiply the left side of Eq. (13) with $\phi_{\vec {k}}^{*}({\bf r})$ and 
integrate over the position,
we find
\begin{eqnarray}
 a_{\vec {k}\uparrow} &=& \sum_{n}(\gamma_{n\uparrow}u_{n, \vec {k}} - \gamma^{\dagger}_{n\downarrow}v_{n,\vec {k}}), \nonumber\\
 a_{\vec {k}\downarrow} &=& \sum_{n}(\gamma_{n\downarrow}u_{n, \vec {k}} + 
\gamma^{\dagger}_{n\uparrow}v_{n,\vec {k}}), 
\end{eqnarray}
where
\begin{eqnarray}
u_{n,\vec {k}} &=& \int \phi^{*}_{\vec {k}}({\bf r}) u_{n}({\bf r})d{\bf r},\nonumber\\
v_{n,\vec {k}} &=& \int \phi^{*}_{\vec {k}}({\bf r})v^{*}_{n}({\bf r})d{\bf r}.
\end{eqnarray}
Let's compare Eqs. (11) and (14). Only if both $u_{n}({\bf r})$ and $v_{n}({\bf r})$ are proportional
to the normal state wavefunction $\phi_{\vec {k}}({\bf r})$, then the transformation (14) becomes  
the same as the BV transformation.  

To understand the physical meaning of the unitary transformation (5), we express $\gamma_{n\uparrow}$ by
the creation and destruction operators for an electron.  
If we multiply $u_{n'}^{*}({\bf r})$ to the first equation of Eq. (13) and 
integrate over the position,
we find
\begin{eqnarray}
\sum_{\vec {k}}u_{n',\vec {k}}^{*}a_{\vec {k}\uparrow} = \sum_{n}\Bigl(\gamma_{n\uparrow}\int u_{n}({\bf r})u_{n'}^{*}({\bf r})d{\bf r} -
   \gamma_{n\downarrow}^{\dagger}\int u_{n'}^{*}({\bf r})v_{n}^{*}({\bf r})d{\bf r}\Bigr).
\end{eqnarray}
By a similar process to the second equation, we also find
\begin{eqnarray}
\sum_{\vec {k}}v_{n',\vec {k}}a_{\vec {k}\downarrow}^{\dagger} = \sum_{n}\Bigl(\gamma_{n\downarrow}^{\dagger}\int u_{n}^{*}({\bf r})
v_{n'}^{*}({\bf r})d{\bf r} + \gamma_{n\uparrow}\int v_{n}({\bf r})v_{n'}^{*}({\bf r})d{\bf r}\Bigr).
\end{eqnarray}
Adding Eqs. (16) and (17) and by the orthogonality condition of eigenfunctions, one can get
\begin{eqnarray}
\gamma_{n'\uparrow} = \sum_{\vec {k}}\bigl( u_{n',\vec {k}}^{*}a_{\vec {k}\uparrow} + v_{n',\vec {k}}a_{\vec {k}\downarrow}^{\dagger}\bigr).
\end{eqnarray}
Similarly it is given 
\begin{eqnarray}
\gamma_{n'\downarrow} = \sum_{\vec {k}}\bigl( u_{n',\vec {k}}^{*}a_{\vec {k}\downarrow} - v_{n',\vec {k}}a_{\vec {k}\uparrow}^{\dagger}\bigr).
\end{eqnarray}
Let's define $\Phi_{n}({\bf r})$ and $\Phi_{\bar n}({\bf r})$ by 
\begin{eqnarray}
\Phi_{n}({\bf r}) = {1\over U_{n}}u_{n}({\bf r}) 
 = {1\over U_{n}} \sum_{\vec {k}}u_{n,\vec {k}}\phi_{\vec {k}}({\bf r}), 
\end{eqnarray}
and 
\begin{eqnarray}
\Phi_{\bar n}({\bf r}) = {-1\over V_{n}}v_{n}^{*}({\bf r}) = 
 {-1\over V_{n}} \sum_{\vec {k}}v_{n,\vec {k}}\phi_{\vec {k}}({\bf r}),
\end{eqnarray}
where 
\begin{eqnarray}
|U_{n}|^{2} = \int u_{n}^{*}({\bf r})u_{n}({\bf r})d{\bf r}, 
\end{eqnarray}
and 
\begin{eqnarray}
|V_{n}|^{2} = \int v_{n}^{*}({\bf r})v_{n}({\bf r})d{\bf r}. 
\end{eqnarray}
Notice that $u_{n}(\bf r)$ and $v_{n}(\bf r)$ were expanded by the
plane wave basis $\phi_{\vec {k}}(\bf r)$ in Eqs. (20) and (21). 
Then we find
\begin{eqnarray}
a_{n\alpha} &=& {1\over U_{n}}\sum_{\vec {k}}u_{n,\vec {k}}^{*}a_{\vec {k}\alpha},\nonumber\\
a_{{\bar n} \alpha} &=& {-1\over V_{n}^{*}}\sum_{\vec {k}}v_{n, \vec {k}}^{*}a_{\vec {k}\alpha}.
\end{eqnarray}
Finally, by substituting Eq. (24) into Eqs. (18) and (19), we obtain
\begin{eqnarray}
\gamma_{n\uparrow} &=& U_{n}a_{n\uparrow} - V_{n}a_{{\bar n}\downarrow}^{\dagger}, \nonumber\\
\gamma_{n\downarrow} &=& U_{n}a_{n\downarrow} + V_{n}a_{{\bar n}\uparrow}^{\dagger}.
\end{eqnarray}

Note that we pair $\Phi_{n}({\bf r})[={1\over U_{n}}u_{n}({\bf r})]\uparrow$ and 
$\Phi_{\bar n}({\bf r})[=-{1\over V_{n}}v^{*}_{n}({\bf r})]\downarrow$
(instead of $\phi_{\vec {k}}({\bf r})\uparrow$ and $\phi_{-\vec {k}}({\bf r})\downarrow$)
by the unitary transformation (5). 
The generated vacuum state is
\begin{eqnarray}
{\tilde \phi_{BdG}} = \prod_{n} (U_{n} + V_{n}a_{n\uparrow}^{\dagger}
         a_{{\bar n}\downarrow}^{\dagger})|0>, 
\end{eqnarray}
instead of the BCS ground state$^{20}$
\begin{eqnarray}
{\tilde \phi_{BCS}} = \prod_{\vec {k}} (u_{\vec {k}} + v_{\vec {k}}a_{\vec {k}\uparrow}^{\dagger}
         a_{-\vec {k}\downarrow}^{\dagger})|0>. 
\end{eqnarray}
Therefore a pairing constraint is necessary for the unitary
transformation (5) to generate the BCS ground state; that is, 
both $u_{n}({\bf r})$ and $v_{n}({\bf r})$ should be proportional to the normal
state wavefunction $\phi_{\vec {k}}(\bf r)$ in order to pair $\vec {k}\uparrow$ and
$-\vec {k}\downarrow$.
For the current-carrying state, we can pair $\vec {k}+{\vec q}\uparrow$ and $-\vec {k}+{\vec q}\downarrow$. 
Then, $u_{n}({\bf r}) = U_{\vec {k}}e^{i(\vec {k}+{\vec q})\cdot \bf r}$ and 
$v^{*}_{n}({\bf r}) = V_{\vec {k}} e^{i(-\vec {k}+{\vec q})\cdot\bf r}$. 
\vskip 1pc
\leftline{\bf 2.3 Pairing Constraint on the Self-consistency Equation}
\vskip 1pc

The pairing constraint should be also supplemented in the self-consistency equation 
$$\Delta({\bf r}) = V\sum_{n}v^{*}_{n}({\bf r})u_{n}({\bf r})(1-2f_{n}).\eqno(10)$$
However, $u_{n}$ and $v_{n}$ are usually expanded as a power series 
in $\Delta$$^{15}$ 
\begin{eqnarray}
u_{n} &=& u_{n}^{o}+u_{n}^{1} + \cdots,\nonumber\\
v_{n} &=& v_{n}^{o} + v_{n}^{1} + \cdots, 
\end{eqnarray}
where
\begin{eqnarray}
u_{n}^{o} &=& \phi_{\vec {k}}, \quad v_{n}^{o}=0 \quad (\xi_{\vec {k}} > 0),\nonumber\\
u_{n}^{o} &=& 0, \quad v_{n}^{o}=\phi_{\vec {k}}^{*} \quad (
\xi_{\vec {k}} < 0), 
\end{eqnarray}
and 
\begin{eqnarray}
u_{n}^{1} &=& \sum_{\vec {k}'}e_{n\vec {k}'}\phi_{\vec {k}'},\nonumber\\
v_{n}^{1} &=& \sum_{\vec {k}'}d_{n\vec {k}'}\phi_{\vec {k}'}^{*}.
\end{eqnarray}
$e_{n\vec {k}'}$ and $d_{n\vec {k}'}$ are given 
\begin{eqnarray}
(|\xi_{n}|-\xi_{\vec {k}'})e_{n\vec {k}'} &=& \int \Delta({\bf r})
\phi_{\vec {k}'}^{*}({\bf r})v_{n}^{o}({\bf r})d{\bf r}\nonumber\\
(|\xi_{n}|+\xi_{\vec {k}'})d_{n\vec {k}'} &=& \int \Delta^{*}({\bf r})
\phi_{\vec {k}'}({\bf r})u_{n}^{o}({\bf r})d{\bf r}, 
\end{eqnarray}
with $\xi_{\vec {k}} = {\hbar^{2}\vec {k}^{2}\over 2m} - E_{F} $.
In that case both $u_{n}$ and $v_{n}$ are the linear combiantion of the
normal state wavefunction, which violates the pairing constraint.
If we substitute Eq. (28) to the self-consistency equation Eq. (10),
we obtain
\begin{eqnarray}
\Delta({\bf r}) = \int K({\bf r},{\bf l})\Delta({\bf l})d{\bf l}, 
\end{eqnarray}
where
\begin{eqnarray}
K({\bf r},{\bf l}) = VT\sum_{\omega}\sum_{\vec {k}\vec {k}'}{\phi_{\vec {k}}^{*}({\bf l})
\phi_{\vec {k}'}^{*}({\bf l})\phi_{\vec {k}}({\bf r})\phi_{\vec {k}'}({\bf r})\over (\xi_{\vec {k}}
- i\omega)(\xi_{\vec {k}'}+i\omega)}.
\end{eqnarray}
The $\omega$'s are $\omega_{n}=(2n+1)\pi T$ for all integer $n$.	
Consequently, this equation should be corrected by a pairing
constraint. Then we find
\begin{eqnarray}
\Delta({\bf r}) = \int K^{c}({\bf r},{\bf l})\Delta({\bf l})d{\bf l}, 
\end{eqnarray}
where
\begin{eqnarray}
K^{c}({\bf r},{\bf l}) = VT\sum_{\omega}\sum_{\vec {k}\vec {k}'}{\phi_{\vec {k}}^{*}({\bf l})
\phi_{\vec {k}'}^{*}({\bf l})\phi_{\vec {k}}({\bf r})\phi_{\vec {k}'}({\bf r})\over (\xi_{\vec {k}}
- i\omega)(\xi_{\vec {k}'}+i\omega)}
\delta_{\vec {k}'=-\vec {k}}.
\end{eqnarray}

Note that Eq. (32) is the Gor'kov's self-consistency 
equation
\begin{eqnarray}
\Delta({\bf r}) = VT\sum_{\omega}\int \Delta({\bf l})G_{\omega}({\bf r,\bf l})
G_{-\omega}({\bf r,\bf l})d\bf l.
\end{eqnarray}
The revised self-consistency equation (34) can be written
\begin{eqnarray}
\Delta({\bf r}) = VT\sum_{\omega}\int \Delta({\bf l})\{G_{\omega}({\bf r,\bf l})
G_{-\omega}({\bf r,\bf l})\}_{p.p.}d{\bf l},
\end{eqnarray}
where p.p. means proper pairing constraint which dictates pairing
between $\vec {k}\uparrow$ and $-\vec {k}\downarrow$.
It was pointed out$^{13}$ that Eq. (36) was obtained from the Anomalous
Green's function $F({\bf r}, {\bf r}')$ which includes the terms violating
the homogeneity constraint $F({\bf r},{\bf r'})=F({\bf r}-{\bf r'})$. 
Accordingly, the same pairing constraint may be obtained from the
homogeneity constraint of the Anomalous Green's function.$^{13}$	

\vskip 1pc
\leftline{\bf 3. Inhomogeneous System: Nonmagnetic impurity case }
\vskip 1pc
\leftline{\bf 3.1 Correct Vacuum State}
\vskip 1pc
In the presence of ordinary impurities, 
Anderson$^{5}$ proposed a BCS type theory which employs time-reversed
scattered state pairs.
The exact scattered states $\psi_{n}$ satisfy the equation
\begin{eqnarray}
[{p^{2}\over 2m} + U({\bf r})]\psi_{n}({\bf r}) = \xi_{n}\psi_{n}({\bf r}),
\end{eqnarray}
where 
\begin{eqnarray}
U({\bf r}) = \sum_{i}V_{o}\delta({\bf r}-{\bf R}_{i}).
\end{eqnarray}
$\{{\bf R}_{i}\}$ are the impurity sites.
Accordingly, the ground state is$^{15}$
\begin{eqnarray}
{\tilde \phi_{Anderson}} = \prod_{n} (u_{n} + v_{n}c_{n\uparrow}^{\dagger}
         c_{{\bar n}\downarrow}^{\dagger})|0>, 
\end{eqnarray}
where $c_{{\bar n}\downarrow}^{\dagger}$ is the creation operator for an electron
in the state $\psi_{n}^{*}({\bf r})|\downarrow>$.

On the other hand, it has been claimed$^{15}$ that the energy is lowered
if we pair states $\tilde \Phi_{n}$ which are better choices than $\psi_{n}$ 
by using the Bogoliubov-de Gennes equations. 
The state $\tilde \Phi_{n}$ is basically a linear combination of the
normal scattered states.
The coupling comes from the pair potential.
However, we show that 
pairing $\tilde \Phi_{n}\uparrow$ and $\tilde \Phi_{\bar n}\downarrow$
leads to the violation of the physical constraint of the system.

The effective Hamiltonian is given
\begin{eqnarray}
H'_{eff} =  \int d{\bf r}\{\sum_{\alpha}\Psi^{\dagger}({\bf r}\alpha)
[{\vec {p}^{2}\over 2m} + U({\bf r})]
\Psi({\bf r}\alpha) + \Delta({\bf r}) \Psi^{\dagger}({\bf r}\uparrow) \Psi^{\dagger}({\bf r}\downarrow)
+ \Delta^{*}({\bf r}) \Psi({\bf r}\downarrow) \Psi({\bf r}\uparrow)\},
\end{eqnarray}
where 
\begin{eqnarray}
\Delta({\bf r}) = -V<\Psi({\bf r}\downarrow)\Psi({\bf r}\uparrow)>.
\end{eqnarray}
As in Sec. 2.1, the unitary transformation
\begin{eqnarray}
\Psi({\bf r}\uparrow) &=& \sum_{n}(\gamma_{n\uparrow}u_{n}({\bf r}) - 
\gamma^{\dagger}_{n\downarrow}v^{*}_{n}({\bf r})), \nonumber\\
\Psi({\bf r}\downarrow) &=& \sum_{n}(\gamma_{n\downarrow}u_{n}({\bf r}) + 
\gamma^{\dagger}_{n\uparrow}v^{*}_{n}({\bf r})), 
\end{eqnarray}
leads to the following Bogoliubov-de Gennes equations
\begin{eqnarray}
\epsilon u({\bf r}) &=& [H_{e}+U({\bf r})]u({\bf r}) + \Delta({\bf r})v({\bf r}), \nonumber\\ 
\epsilon v({\bf r}) &=& -[H_{e}^{*}+U({\bf r})]v({\bf r}) + \Delta^{*}({\bf r})u({\bf r}). 
\end{eqnarray}

To find the vacuum state for $\gamma$ particles, we expand the field
operator by the scattered states:
\begin{eqnarray}
\Psi({\bf r}\alpha) = \sum_{n} \psi_{n}({\bf r}) c_{n\alpha}.
\end{eqnarray}
Then it can be shown 
\begin{eqnarray}
\gamma_{n\uparrow} &=&\sum_{n'}\bigl( u_{n,n'}^{*}c_{n'\uparrow} + v_{n,n'}c_{n'\downarrow}^{\dagger}\bigr),\nonumber\\
\gamma_{n\downarrow} &=&\sum_{n'}\bigl( u_{n,n'}^{*}c_{n'\downarrow} - v_{n,n'}c_{n'\uparrow}^{\dagger}\bigr),
\end{eqnarray}
where
\begin{eqnarray}
u_{n,n'} &=& \int \psi^{*}_{n'}({\bf r}) u_{n}({\bf r})d{\bf r},\nonumber\\
v_{n,n'} &=& \int \psi^{*}_{n'}({\bf r})v^{*}_{n}({\bf r})d{\bf r}. 
\end{eqnarray}
For the states ${\tilde \Phi}_{n}$ and ${\tilde \Phi}_{\bar n}$
defined by
\begin{eqnarray}
{\tilde \Phi}_{n}({\bf r}) &=&{1\over U_{n}}u_{n}({\bf r})
 ={1\over U_{n}}\sum_{n'}u_{n,n'}\psi_{n'}({\bf r}),\nonumber\\
{\tilde \Phi}_{\bar n}({\bf r}) &=&{-1\over V_{n}}v^{*}_{n}({\bf r})
={-1\over V_{n}}\sum_{n'}v_{n,n'}\psi_{n'}({\bf r}),
\end{eqnarray}
it is given 
\begin{eqnarray}
b_{n\alpha} &=& {1\over U_{n}}\sum_{n'}u_{n,n'}^{*}c_{n'\alpha},\nonumber\\
b_{{\bar n} \alpha} &=& {-1\over V_{n}^{*}}\sum_{n'}v_{n, n'}^{*}c_{n'\alpha}.
\end{eqnarray}
$b_{{\bar n}\alpha}$ is the destruction operator for an
electron in the state ${\tilde \Phi}_{n}\alpha$.

Finally, we obtain 
\begin{eqnarray}
\gamma_{n\uparrow} &=& U_{n}b_{n\uparrow} - V_{n}b_{{\bar n}\downarrow}^{\dagger}, \nonumber\\
\gamma_{n\downarrow} &=& U_{n}b_{n\downarrow} + V_{n}b_{{\bar n}\uparrow}^{\dagger},
\end{eqnarray}
and
\begin{eqnarray}
{\tilde \phi_{BdG}} = \prod_{n} (U_{n} + V_{n}b_{n\uparrow}^{\dagger}
         b_{{\bar n}\downarrow}^{\dagger})|0>. 
\end{eqnarray}
Note that the Bogoliubov-de Gennes equations, Eq.(44) correspond to the 
vacuum state where ${\tilde \Phi}_{n}({\bf r})[={1\over U_{n}}u_{n}({\bf r})]\uparrow$ 
and
 ${\tilde \Phi}_{\bar n}({\bf r})[=-{1\over V_{n}}v^{*}_{n}({\bf r})]\downarrow$ (instead of $\psi_{n}({\bf r})\uparrow$
and $\psi_{\bar n}({\bf r})\downarrow$) are paired.

Now we must decide which is the correct ground state in the presence of 
impurities.
Above all, the correct ground state should satisfy the
physical constraint of the system.
If we average over the impurity positions, the system becomes homogeneous.
Consequently, the pair potential should be a constant after the
impurity average, i.e.,
\begin{eqnarray}
\overline{\Delta({\bf r})}^{imp} = \rm constant. 
\end{eqnarray}
$\bar{\ }\bar{\ }^{imp}$ means an average over impurity positions
$\vec R_{i}$.

Let's first check the constraint for the state $ {\tilde \phi_{BdG}}$. 
Eq. (42) leads to the pair potential at $T=0$,
\begin{eqnarray}
\Delta({\bf r}) &=& V\sum_{n}v^{*}_{n}({\bf r})u_{n}({\bf r})\nonumber\\
 &=& -V\sum_{n}U_{n}V_{n}{\tilde \Phi}_{n}({\bf r}) {\tilde \Phi}_{\bar n}({\bf r}).
\end{eqnarray}
In terms of the scattered states, it is rewritten   
\begin{eqnarray}
\Delta({\bf r}) = -V\sum_{n} \sum_{m,m'}u_{n,m}\psi_{m}({\bf r})
v_{n,m'}\psi_{m'}({\bf r}).
\end{eqnarray}
Subsequently it is given
\begin{eqnarray}
\overline{\Delta({\bf r})}^{imp} \sim 
\overline {\psi_{m}({\bf r})\psi_{m'}({\bf r})}^{imp}.
\end{eqnarray}
Note that
\begin{eqnarray}
\overline{\psi_{m(\vec {k})\uparrow}({\bf r})\psi_{m'(\vec {k}')\downarrow}({\bf r})}^{imp} 
 &=& e^{i(\vec {k}+\vec {k}')\cdot{\bf r}}[1 + V_{o}^{2}
\sum_{\vec {q},i}{1\over 
(\epsilon_{\vec {k}} - \epsilon_{\vec {k}+\vec {q}}) (\epsilon_{\vec {k}'} - \epsilon_{\vec {k}'-\vec {q}})}
+ \cdots ]\nonumber\\
 &\not=& \rm constant,
\end{eqnarray}
and
\begin{eqnarray}
\overline{\psi_{m(\vec {k})\uparrow}({\bf r})\psi_{{\bar m}(-\vec {k})\downarrow}({\bf r})}^{imp} 
 &=& [1 + V_{o}^{2}
\sum_{\vec {q},i}{1\over 
(\epsilon_{\vec {k}} - \epsilon_{\vec {k}+\vec {q}})^{2}} 
+ \cdots ]\nonumber\\
 &=& \rm constant .
\end{eqnarray}
In deriving these relations, we used the scattered states $\psi_{m(\vec {k})}(\bf r)$,
\begin{eqnarray}
\psi_{m(\vec {k})}({\bf r}) = e^{i\vec {k}\cdot{\bf r}} + \sum_{\vec {q}}{V_{o}\over \epsilon_{\vec {k}} -
\epsilon_{\vec {k}+\vec {q}}}[\sum_{i}e^{-i\vec {q}\cdot{\vec R_{i}}}]e^{i(\vec {k}+\vec {q})\cdot {\bf r}}
+  \cdots . 
\end{eqnarray}
Consequently the state ${\tilde \Phi}_{BdG}$ gives the `averaged' pair potential
\begin{eqnarray}
\overline{\Delta({\bf r})}^{imp} \not= \rm constant, 
\end{eqnarray}
which violates the physical constraint of the system.

From Eq. (57), it is clear that the state 
${\tilde \Phi}_{Anderson}$ gives the correct `averaged' pair potential
which satisfy the physical constraint of the system.
Therefore the correct ground state is  
${\tilde \Phi}_{Anderson}$. 
To obtain 
${\tilde \Phi}_{Anderson}$ 
from the Bogoliubov-de Gennes equations,
we need a pairing constraint:
\begin{eqnarray}
{\tilde \phi_{BdG}} = {\tilde \phi_{Anderson}},
\end{eqnarray}
which gives
\begin{eqnarray}
u_{n}({\bf r}) &\propto& \psi_{n}({\bf r}),\nonumber\\
v_{n}^{*}({\bf r}) &\propto& \psi_{\bar n}({\bf r}).
\end{eqnarray}
\vskip 1pc
\leftline{\bf 3.2 Localization Correction}
\vskip 1pc
It has been a common practice to assume a constant pair
potential to derive Anderson's theorem.$^{15,21,22}$
For the constant pair potential $\Delta_{o}$, we can also choose 
\begin{eqnarray}
u({\bf r})&=&\psi_{n}({\bf r})u_{n},\nonumber\\
             v({\bf r})&=&\psi_{n}({\bf r})v_{n}.
\end{eqnarray}
Then Eq. (44), or Eq. (10) leads to 
\begin{eqnarray}
\Delta_{o} = V\sum_{n}u_{n}v_{n}\psi_{n}({\bf r})\psi_{n}^{*}({\bf r})
(1-2f_{n}).
\end{eqnarray}
To be consistent, the right hand side is substitituted by the
impurity average of the square of wavefunction, i.e.,
\begin{eqnarray}
\Delta_{o}=V\overline{N_{o}}\int  d\xi_{n}u_{n}v_{n}(1-2f_{n}),
\end{eqnarray}
where 
\begin{eqnarray}
\overline{N_{o}}=\sum_{n}\delta(\xi_{n})
\overline{\psi_{n}({\bf r})\psi_{n}^{*}({\bf r})}^{imp}.
\end{eqnarray}
Eq. (64) is the same form as that of the homogeneous case.
Consequently, the transition temperature doesn't change in
the presence of nonmagnetic impurities, which is called by Anderson's 
theorem.

However, we should not assume a constant pair potential to derive Eq. (62),
which is just Anderson's pairing constraint.
Then, we obtain the self-consistency equation 
\begin{eqnarray}
\Delta({\bf r}) = V\sum_{n}u_{n}v_{n}\psi_{n}({\bf r})\psi_{n}^{*}({\bf r})
(1-2f_{n}),
\end{eqnarray}
instead of Eq. (63) which is inconsistent.
Multiplying the both sides of Eq. (66) by $\psi_{m}^{*}({\bf r})\psi_{\bar m}^{*}({\bf r})$,
one finds that
\begin{eqnarray}
\Delta_{m} = \sum_{n}V_{mn}u_{n}v_{n}(1-2f_{n}),
\end{eqnarray}
where
\begin{eqnarray}
V_{mn}=V\int \psi_{m}^{*}({\bf r})\psi_{\bar m}^{*}({\bf r})\psi_{n}({\bf r})\psi_{\bar n}({\bf r})d{\bf r}.
\end{eqnarray}
And it is given
\begin{eqnarray}
u_{n}&=&{1\over 2}(1+{\xi_{n}\over \sqrt{\xi_{n}^{2}+
\Delta_{n}^{2}}}),\nonumber\\
v_{n}&=&{1\over 2}(1-{\xi_{n}\over \sqrt{\xi_{n}^{2}+
\Delta_{n}^{2}}}).
\end{eqnarray}
Comparing Eqs. (67) and (64), we find that Anderson's theorem is valid 
only when $V_{mn}$  is not much different from $V$, which is the  effective
interaction without impurities.
It was pointed out that this quantity is almost the same as $V$ up to the first
order of the impurity  concentration.$^{1,2}$

Now it is clear that localization correction is important in Eq. (68).
For the strongly localized states, the effective interaction is
exponentially small,$^{1,6}$ like the conductance.$^{23,24}$
It is, then, expected that the same weak localization correction terms occur
both in the conductance and the effective interaction.
Recently weak localization correction to the phonon-mediated
interaction was reported.$^{13}$
The results are the following:
\begin{eqnarray}
V_{nn'}^{3d} \cong -V[1-{1\over (k_{F}\ell)^{2}}(1-{\ell\over L})],
\end{eqnarray}
\begin{eqnarray}
V_{nn'}^{2d} \cong  
 -V[1-{2\over \pi k_{F}\ell}ln(L/\ell)],
\end{eqnarray}
\begin{eqnarray}
V_{nn'}^{1d} \cong  
 -V[1-{1\over (\pi k_{F}a)^{2}}(L/\ell-1)],
\end{eqnarray}
where $\ell$ and $L$ are the elastic and inelastic mean free paths and
$a$ is the radius of the wire.
For thin films, the empirical formula is given$^{25}$	
\begin{eqnarray}
{T_{co}-T_{c}\over T_{co}}\propto R_{sq},
\end{eqnarray}
where $T_{co}$ is the unperturbed value of $T_{c}$ and 
$R_{sq}$ is the resistance of a square sample. 
Notice that this formula is obtained if we substitute 
Eq. (71) into the BCS gap equation.
More details will be published elsewhere.

\vskip 1pc
\leftline {\bf 4. Conclusion}
\vskip 1pc

It is shown that the Bogoliubov-de Gennes equations need a pairing
constraint to obtain a correct vacuum state by the corresponding transformation.
The constraint is the same as that obtained from the physical constraint of 
the Anomalous Green's function. 

\vskip 1pc
\leftline {\bf Acknowledgments}
\vskip 1pc

I am grateful to Professor A. W. Overhauser for discussions.  
The early version of the paper was circulated in the U.S.A., Japan,
and Korea from early 1995 to mid 1995.
This work was supported by the National Science Foundation, Materials Theory
Program.
\vfill\eject
\centerline      {\bf REFERENCES} 
\vskip 1pt\hb
1. Yong-Jihn Kim and A. W. Overhauser, Phys. Rev. B{\bf 47}, 8025 (1993).\hb
2. Yong-Jihn Kim and A. W. Overhauser, Phys. Rev. B{\bf 49}, 12339 (1994).\hb
3. A. A. Abrikosov and L. P. Gor'kov, Sov. Phys. JETP {\bf 12}, 1243 (1961).\hb
4. A. A. Abrikosov and L. P. Gor'kov, Phys. Rev. B{\bf 49}, 12337 (1994).\hb
5. P. W. Anderson, J. Phys. Chem. Solids {\bf 11}, 26 (1959).\hb
6. Yong-Jihn Kim, Mod. Phys. Lett. B{\bf 10}, 353 (1996).\hb
7. Yong-Jihn Kim and A. W. Overhauser, Phys. Rev. B{\bf 49}, 15779 (1994).\hb
8. M. F. Merriam, S. H. Liu, and D. P. Seraphim, Phys. Rev. {\bf 136}, A17 (1964).\hb
9. G. Boato, M. Bugo, and C. Rizzuto, Phys. Rev. {\bf 148}, 353 (1966).\hb
10. G. Boato and C. Rizzuto, (to be published), (referenced in A. J. Heeger, (1969), in

Solid State Physics {\bf 23}, eds. F. Seitz, D. Turnbull and H. Ehrenreich 
(Academic

 Press, New York), p. 409)  \hb
11. W. Bauriedl and G. Heim, Z. Phys. B{\bf 26}, 29 (1977); M. Hitzfeld and G. Heim, Sol. 

Sta. Com. {\bf 29}, 93 (1979).\hb  
12. A. Hofmann, W. Bauriedl, and P. Ziemann, Z. Phys. B{\bf 46}, 117 (1982).\hb
13. Yong-Jihn Kim, Mod. Phys. Lett. B{\bf 10}, 555 (1996).\hb
14. L. P. Gor'kov, Sov. Phys. JETP {\bf 7}, 505 (1958); {\sl ibid.} {\bf 9}, 1364(1959).\hb
15. P. G. de Gennes, {\sl Superconductivity of Metals and Alloys} (Benjamin, New York, 1966).\hb 
16. N. N. Bogoliubov, JETP USSR {\bf 34}, 58 (1958).\hb
17. J. G. Valatin, Nuovo Cim. {\bf 7}, 843 (1958).\hb
18. C. Bloch and A. Messiah, Nucl. Phys. {\bf 39}, 95 (1962).\hb
19. Fan Hong-Yi and J. VanderLinde, J. Phys. A: Math. Gen. {\bf 23}, L1113 (1990).\hb
20. K. Yosida, Phys. Rev. {\bf 111}, 1255 (1958).\hb
21. P. G. de Gennes and G. Sarma, J. Appl. Phys. {\bf 34}, 1380 (1963).\hb
22. M. Ma and P. A. Lee, Phys. Rev. B{\bf 32}, 5658 (1985).\hb
23. N. F. Mott and M. Kaveh, Adv. Phys. {\bf 34}, 329 (1985).\hb
24. P. A. Lee and T. V. Ramakrishnan, Rev. Mod. Phys. {\bf 57}, 287 (1985).\hb
25. B. I. Belevtsev, Sov. Phys. Usp. {\bf 33}, 36 (1990).\hb
\end{document}